\documentclass{emulateapj}

\usepackage{amsmath}

\newcommand{\about}{$\sim\!\!$~}

\newcommand{\be}{\begin{displaymath}}
\newcommand{\ee}{\end{displaymath}}

\def\lsim{\hbox{\rlap{\raise 0.425ex\hbox{$<$}}\lower 0.65ex\hbox{$\sim$}}}
\def\gsim{\hbox{\rlap{\raise 0.425ex\hbox{$>$}}\lower 0.65ex\hbox{$\sim$}}}
\def\arcmin{\hbox{$^\prime$}}

\def\arcdeg{\mbox{$^\circ$}}

\newcommand{\kms}{km~s$^{-1}$}

\shorttitle{The Very Young Type Ia SN 2013dy}
\shortauthors{Zheng et al.}

\begin{document}

\title{The Very Young Type Ia Supernova 2013dy: Discovery, and Strong Carbon Absorption in Early-Time Spectra}

\author{WeiKang Zheng\altaffilmark{1,2},
Jeffrey M. Silverman\altaffilmark{3,4},
Alexei V. Filippenko\altaffilmark{1},
Daniel Kasen\altaffilmark{5,6},
Peter E. Nugent\altaffilmark{5,1},
Melissa Graham\altaffilmark{1,7,8},
Xiaofeng Wang\altaffilmark{9},
Stefano Valenti\altaffilmark{7,8},
Fabrizio Ciabattari\altaffilmark{10},
Patrick L. Kelly\altaffilmark{1},
Ori D. Fox\altaffilmark{1},
Isaac Shivvers\altaffilmark{1},
Kelsey I. Clubb\altaffilmark{1}, 
S. Bradley Cenko\altaffilmark{11},
Dave Balam\altaffilmark{12},
D. Andrew Howell\altaffilmark{7,8},
Eric Hsiao\altaffilmark{13},
Weidong Li\altaffilmark{1,14},
G. Howie Marion\altaffilmark{3,15},
David Sand\altaffilmark{16},
Jozsef Vinko\altaffilmark{17,3},
J. Craig Wheeler\altaffilmark{3}, and
JuJia Zhang\altaffilmark{18,19}
}

\altaffiltext{1}{Department of Astronomy, University of California, Berkeley, CA 94720-3411, USA.}
\altaffiltext{2}{e-mail: zwk@astro.berkeley.edu .}
\altaffiltext{3}{Department of Astronomy, University of Texas, Austin, TX 78712, USA.}
\altaffiltext{4}{NSF Astronomy and Astrophysics Postdoctoral Fellow.}
\altaffiltext{5}{Lawrence Berkeley National Laboratory, Berkeley, California 94720, USA.}
\altaffiltext{6}{Department of Physics, University of California, Berkeley, 94720, USA.}
\altaffiltext{7}{Las Cumbres Observatory Global Telescope Network, 6740 Cortona Drive, Suite 102, Santa Barbara, CA 93117, USA.}
\altaffiltext{8}{Department of Physics, Broida Hall, University of California, Santa Barbara, CA 93106, USA.}
\altaffiltext{9}{Department of Physics, Tsinghua University, Beijing 100084, China.}
\altaffiltext{10}{Monte Agliale Observatory, Borgo a Mozzano, Lucca, 55023 Italy.}
\altaffiltext{11}{Astrophysics Science Division, NASA Goddard Space Flight Center, Mail Code 661, Greenbelt, MD 20771, USA.}
\altaffiltext{12}{Dominion Astrophysical Observatory, National Research Council of Canada.}
\altaffiltext{13}{Carnegie Observatories, Las Campanas Observatory, Colina El Pino, Casilla 601, Chile.}
\altaffiltext{14}{Deceased 2011 December 11.}
\altaffiltext{15}{Harvard-Smithsonian Center for Astrophysics, 60 Garden St., Cambridge, MA 02138, USA.}
\altaffiltext{16}{Physics Department, Texas Tech University, Lubbock, TX 79409, USA.}
\altaffiltext{17}{Department of Optics and Quantum Electronics, University of Szeged, D\'{o}m t\'{e}r 9, 6720 Szeged, Hungary.}
\altaffiltext{18}{Yunan Astronomical Observatory, Chinese Academy of Sciences, 650011, Yunnan, China.}
\altaffiltext{19}{Key Laboratory for the Structure and Evolution of Celestial Objects, Chinese Academy of Sciences, Kunming 650011, China.}

\begin{abstract}
The Type~Ia supernova (SN~Ia) 2013dy in NGC~7250 ($d \approx$ 13.7\,Mpc)
was discovered by the Lick Observatory Supernova Search. Combined with 
a prediscovery detection by the Italian Supernova Search Project, we are 
able to constrain the first-light time of SN~2013dy to be only 
$0.10\pm0.05$\,d ($2.4\pm1.2$\,hr) before the first detection.
This makes SN~2013dy the earliest known detection of an SN~Ia.
We infer an upper limit on the radius of the progenitor star of 
$R_0 \lesssim 0.25\,{\rm R}_\sun$, consistent with that of a white dwarf.
The light curve exhibits a broken power law with exponents of 0.88
and then 1.80. A spectrum taken 1.63\,d after first light
reveals a \ion{C}{2} absorption line comparable in strength to \ion{Si}{2}. 
This is the strongest \ion{C}{2} feature ever detected in a normal SN~Ia, 
suggesting that the progenitor star had significant unburned material. 
The \ion{C}{2} line in SN 2013dy weakens rapidly and is undetected in a
spectrum 7 days later, indicating that \ion{C}{2} is detectable for
only a very short time in some SNe~Ia. SN~2013dy reached a $B$-band 
maximum of $M_B = -18.72 \pm 0.03$\,mag \about17.7~d after first light.
\end{abstract}


%%%%%%%%%%%%%%%%%%%%%%%%%%%%%%%%
%%  Section 1:  Introduction  %%
%%%%%%%%%%%%%%%%%%%%%%%%%%%%%%%%

\section{Introduction}\label{s:intro}

Type~Ia supernovae (SNe~Ia) are used as calibratable candles with
many important applications, including measurements of the expansion rate of the Universe
(Riess et al. 1998; Perlmutter et al. 1999). 
However, the understanding of their progenitor systems and explosion mechanisms remains substantially incomplete.
It is thought that SNe~Ia are the product of the thermonuclear explosions of C/O white
dwarfs (Hoyle \& Fowler 1960; Colgate \& McKee 1969; see Hillebrandt \& Niemeyer 2000
for a review), but very early discovery and detailed follow-up observations are essential for learning about the nature of the progenitor evolution and the nature of the explosion process.
Recent examples of well-studied SNe~Ia include SN~2009ig (Foley et al. 2012), 
SN~2011fe (Nugent et al. 2011; Li et al. 2011), 
and SN~2012cg (Silverman et al. 2012a); like 
SN~2013dy, they were discovered shortly after exploding.

Early discovery and identification give us the opportunity to obtain spectra when the SNe are still very young,
yielding more insight into the composition of the SN blastwave (especially the outer layers) and its progenitor star.
For example, while O is often seen (can be from both unburned material and  a product of C burning), spectroscopic C is 
much more rare. In particular, strong C features have been seen only
in a few ``super-Chandrasekhar mass'' SNe~Ia: SNLS-03D3bb (SN~2003fg;
Howell et al. 2006), SN~2006gz (Hicken et al. 2007), SN~2007if
(Scalzo et al. 2010), and SN~2009dc (Yamanaka et al. 2009; Silverman et al. 2011; Taubenberger et al. 2011).
Though often detectable in normal SNe~Ia, C lines are usually not strong (e.g., Patat et al. 1996; Garavini et al.
2005; Nugent et al. 2011; Silverman et al. 2012b). 

Here we present our observations and analysis of SN~2013dy, detected
merely 0.10\,d after first light.  An early spectrum (1.63\,d) exhibits
an unusually strong absorption feature $\sim 245$\,\AA\ redder than
\ion{Si}{2} $\lambda$6355, very likely produced by \ion{C}{2}.

%%%%%%%%%%%%%%%%%%%%%%%%%%%%%%%%%%%%%%%%%%%%%%%%
%%  Section 2:  Discovery %%
%%%%%%%%%%%%%%%%%%%%%%%%%%%%%%%%%%%%%%%%%%%%%%%%

\section{Discovery and Observations}\label{s:discovery}

The field of NGC~7250 has been observed by the 0.76\,m 
Katzman Automatic Imaging Telescope (KAIT)
more than 600 times over the past 15\,yr as part 
of the Lick Observatory Supernova Search (LOSS; Filippenko et al. 2001).
In early 2011, the LOSS search strategy was modified to monitor fewer
galaxies at a more rapid cadence with the objective of promptly
identifying very young SNe (hours to days after explosion).
The new software autonomously prompts KAIT to obtain a sequence of $U$, $B$, $V$, and 
unfiltered (roughly $R$) images when a new transient is discovered,
usually only minutes after the discovery images were taken. One of the first successful discoveries using
this technique was SN~2012cg (Silverman et al. 2012a), followed by several others
(e.g., SN~2013ab, Blanchard et al. 2013; SN~2013dh, Kumar et al. 2013).
Although multi-band follow-up photometry was not autonomously triggered
for SN~2013dy on the night of discovery, it was triggered
two days later.
The trigger was not activated the first night because the SN was quite faint and
multiple other (spurious) candidates were found in the discovery image.
However, the autonomous trigger activated by the second KAIT image demonstrates that the software triggering
capability functions well.

\begin{figure}[!hbp]
\centering
\includegraphics[width=.235\textwidth]{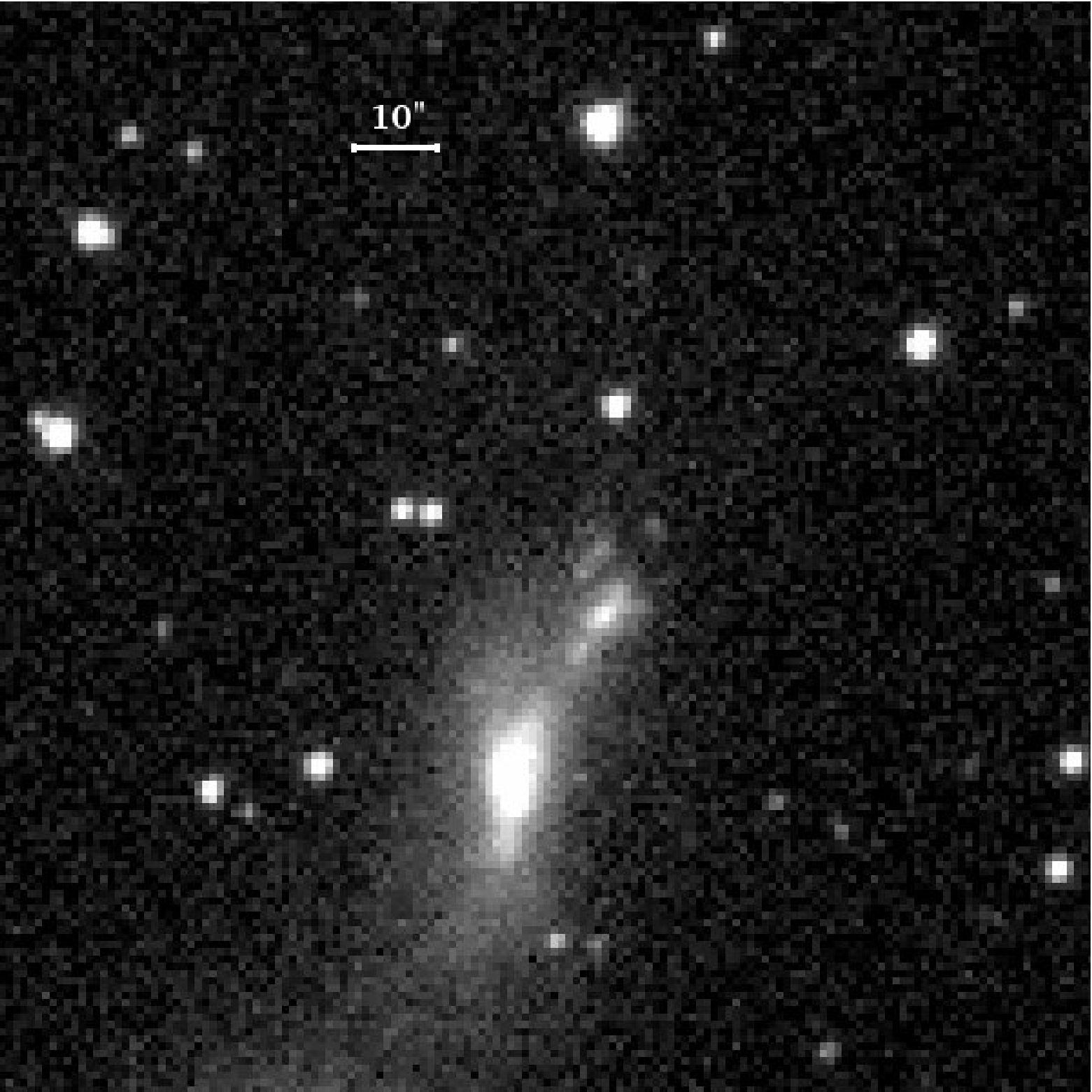}
\includegraphics[width=.235\textwidth]{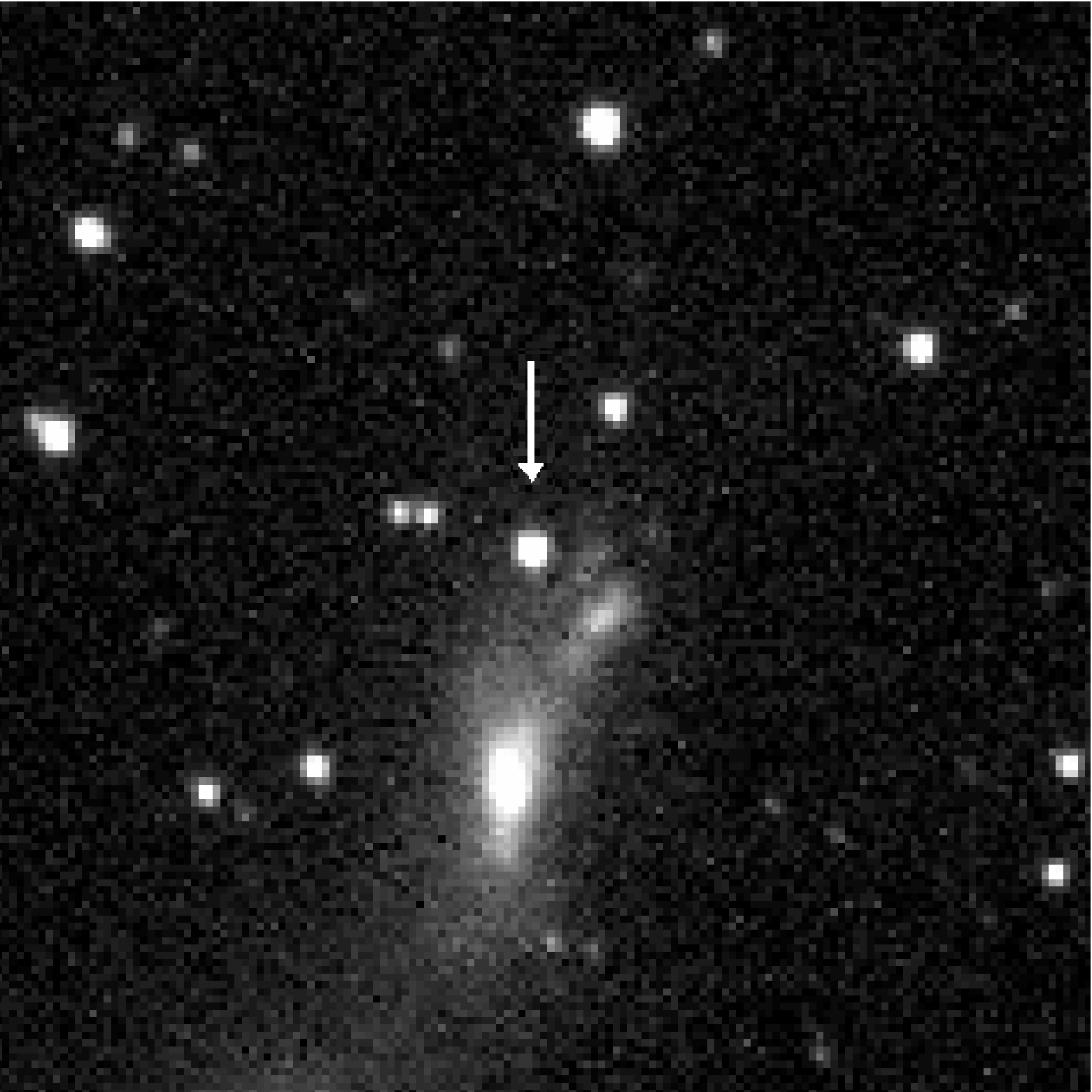}
\includegraphics[width=.475\textwidth]{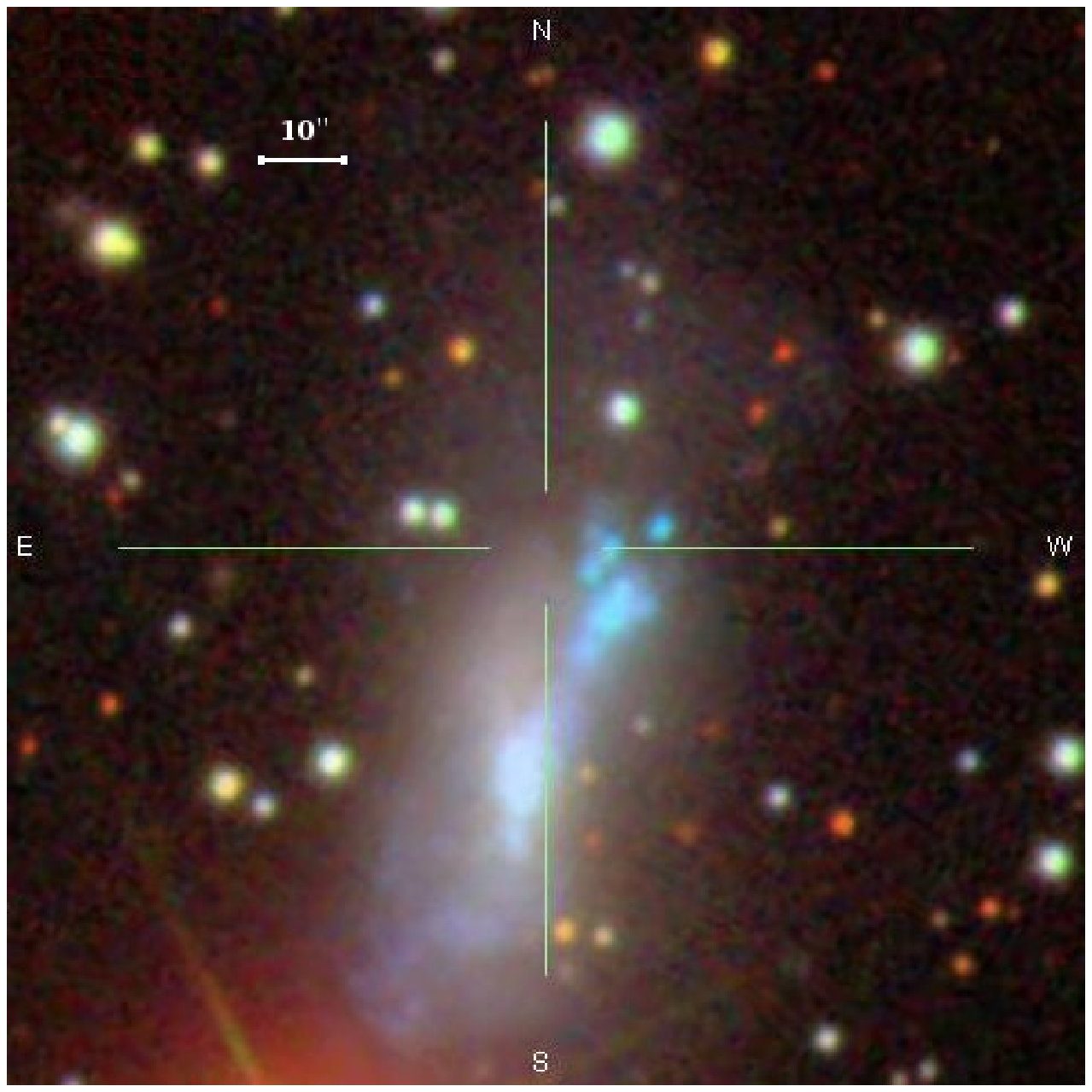}
\caption{Top left: KAIT unfiltered template image. Top right: KAIT
         unfiltered image with the SN indicated by the arrow. Bottom:
	 SDSS color composite of the field around SN~2013dy (position
         marked with crosshairs); the nucleus of
	 NGC~7250 is to the south ($26\farcs5$ away), and a blue 
         star-forming region is to the southwest ($\sim10\farcs7$ away).}
\label{FC_SDSS}
\end{figure}

SN~2013dy was discovered (Casper et al. 2013) in an 18~s unfiltered KAIT 
image taken at 10:55:30 on 2013~July~10 (UT dates are used throughout)
at $R=17.19\pm0.05$\,mag.
We measure its J2000.0 coordinates to be 
$\alpha=22^{\mathrm{h}}18^{\mathrm{m}}17\farcs603$, 
$\delta=+40^{\circ}34\arcmin09\farcs54$, 
with an uncertainty of $0\farcs15$ in each coordinate.
Figure \ref{FC_SDSS} shows KAIT and the Sloan Digital Sky Survey (SDSS) finding chart near the SN location.
SN~2013dy is $2\farcs3$ west and $26\farcs4$ north of
the nucleus of the host galaxy NGC~7250, at a
distance of $13.7\pm3.0$\,Mpc (calculated from the Tully-Fisher relation; Tully et al. 2009),
which gives the SN a projected distance of \about1.76\,kpc from the nucleus.
We note that there is a bright, blue region about $8\farcs7$ west and $6\farcs4$
south of the SN (projected distance $\sim0.71$\,kpc), which may be a star-forming
region or merger (LEDA~214816; Paturel et al. 2000).
It has been recently reported that the observed differences among SNe~Ia
may be tied to their birthplace environments (e.g., Kelly et al. 2010; Wang et al. 2013).
However, it is unclear whether SN~2013dy has any connection with this star-forming region.

We obtained KAIT multi-band images almost every night for the following $\sim3$ weeks,
and they were reduced using our
image-reduction pipeline (Ganeshalingam et al. 2010). Point-spread function
photometry was then
obtained using DAOPHOT (Stetson 1987) from the IDL Astronomy User's
Library\footnote{http://idlastro.gsfc.nasa.gov/.}.
The SN instrumental magnitudes are calibrated to 
local SDSS standards transformed into the
Landolt system
\footnote{http://www.sdss.org/dr7/algorithms/
sdssUBVRITransform.html\#Lupton2005.}.
We applied an image-subtraction procedure to remove host-galaxy light from only the unfiltered images,
because multi-band images without the SN are not yet available. However, KAIT has a relatively small
pixel scale ($0\farcs78$ pixel$^{-1}$), and the host background is quite uniform and faint 
in the KAIT images, so we believe that the contribution from the host galaxy is
minor in all bands, especially considering the brightness of the SN. Comparisons of the
subtracted and not subtracted unfiltered images yield nearly identical
results (differences of $\sim0.1$\,mag or less).

Interestingly, an unfiltered prediscovery detection of SN~2013dy was 
obtained at 02:04:11 July 10 (Casper et al. 2013)
with the 0.5\,m reflector at Monte Agliale Observatory
as part of the Italian Supernova Search Project (ISSP).
Additional confirmation images were taken on July 11 and 26.
We have reprocessed the original images as part of this study.
Owing to the relatively large pixel scale ($2\farcs32$ pixel$^{-1}$), the SN is blended with host-galaxy light.
Using a template image taken on 2011 August 4, we performed
the same subtraction method as for the KAIT unfiltered images.
We then obtained photometry with an aperture of radius 1.5\,pixels, 
a reasonable size given the seeing and large pixel scale.

Additional multi-band photometry in Johnson-Cousins {\it BVRI} was obtained with the Las Cumbres
Observatory Global Telescope (LCOGT) network of robotic 1.0\,m telescopes (Brown et al.
2013). The LCOGT instrumental magnitudes are calibrated to local SDSS standards,
transformed to 
{\it BVRI}\footnote{http://www.sdss.org/dr7/algorithms/
sdssUBVRITransform.html\#Jester2005.}.

Optical spectra of SN~2013dy were obtained on 8 different nights with
DEIMOS (Faber et al. 2003) on the Keck~II telescope (1.63\,d),
the 1.82\,m Plaskett Telescope of the National Research Council of Canada (3.30\,d),
YFOSC on the 2.4\,m telescope at LiJiang Gaomeigu Station of YNAO (4.76\,d),
the Kast double spectrograph (Miller \& Stone 1993) on the Shane 3\,m telescope at Lick Observatory (5.43\,d),
the FLOYDS robotic spectrograph (Sand et al., in prep.) on the LCOGT 2.0\,m Faulkes Telescope North on Haleakala, Hawaii 
(7.50, 8.57, 10.57\,d), and the Marcario Low-Resolution Spectrograph (LRS; Hill et al. 1998)
on the 9.2\,m Hobby-Eberly Telescope (HET) at McDonald Observatory (11.27\,d).
Data were reduced following standard techniques for CCD processing and spectrum extraction using IRAF.
The spectra were flux calibrated through observations of appropriate 
spectrophotometric standard stars.

%%%%%%%%%%%%%%%%%%%%%%%%%%%
%%  Section 3:  Analysis & Results %%
%%%%%%%%%%%%%%%%%%%%%%%%%%%

\section{Analysis and Results}\label{s:analysis}

\subsection{Light Curves}\label{ss:lightcurves}

\begin{figure}[!hbp]
\centering
\includegraphics[width=.49\textwidth]{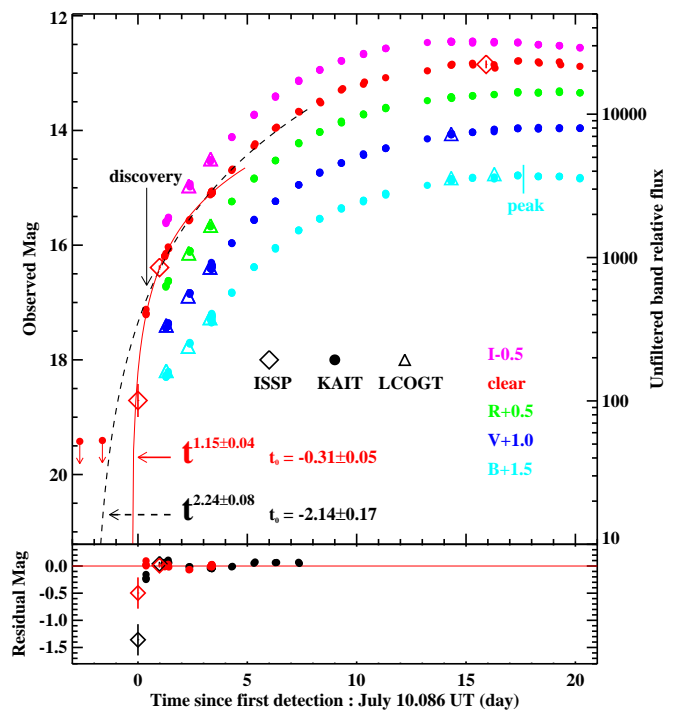}
\caption{Multi-band light curves of SN~2013dy (top panel). Statistical 
  errors are smaller than
  data points if not shown, and include any subtraction error.
  The solid red line is the $t^{1.15}$ fit
  for combined KAIT and ISSP unfiltered fluxes before July 14 (with 
  an inferred first-light time of $-0.31$\,d), while the dashed black 
  line is the $t^{2.24}$ fit only for KAIT data before July 18 (with 
  an inferred first-light time of $-2.14$\,d). The residuals are shown in
  the bottom panel with the same color. This result indicates a varying 
  (broken) power law of the early rising light curve, which we adopt; 
  see Figure \ref{figure_lc_brokenPL}.}
\label{Fig_lcfit}
\end{figure}

Figure \ref{Fig_lcfit} shows our {\it BVRI} and unfiltered light curves of SN~2013dy.
Applying a low-order polynomial fit, we find that SN~2013dy reached a 
$B$-band peak magnitude of $13.28\pm0.03$ on 2013 July $27.71\pm0.30$,
\about17.7\,d after first light.
Assuming $E(B-V)_\textrm{MW} = 0.15$\,mag (Schlegel et al. 1998),
$E(B-V)_\textrm{host} = 0.15$\,mag (see below),
and $d = 13.7$\,Mpc (Tully et al. 2009), this implies $M_B = -18.72 \pm 0.03$ (statistical only) mag,
which is \about0.5\,mag dimmer than the typical SN~Ia, but still within
the range of a ``normal'' SN~Ia. 
The unfiltered band reached a peak of $12.81\pm0.03$\,mag, which means our first detection of the
SN from the ISSP image (18.71\,mag, with a limiting magnitude of $\sim19.5$)
was taken when the SN was at only $\sim0.43$\% of its peak brightness.

%1/230th as bright.

%Extinction
%Landolt U (0.34)  0.831   
%Landolt B (0.44)  0.660   
%Landolt V (0.54)  0.507   
%Landolt R (0.65)  0.409   
%Landolt I (0.81)  0.297   
%SDSS    u (0.35)  0.789   
%SDSS    g (0.49)  0.580   
%SDSS    r (0.63)  0.421   
%SDSS    i (0.78)  0.319   
%SDSS    z (0.93)  0.226   
%UKIRT   J (1.27)  0.138   
%UKIRT   H (1.67)  0.088   
%UKIRT   K (2.22)  0.056   
%UKIRT   L'(3.81)  0.023

%\begin{figure}[!]

%\begin{figure}[!]
\begin{figure}[!hbp]
\centering
\includegraphics[width=.49\textwidth]{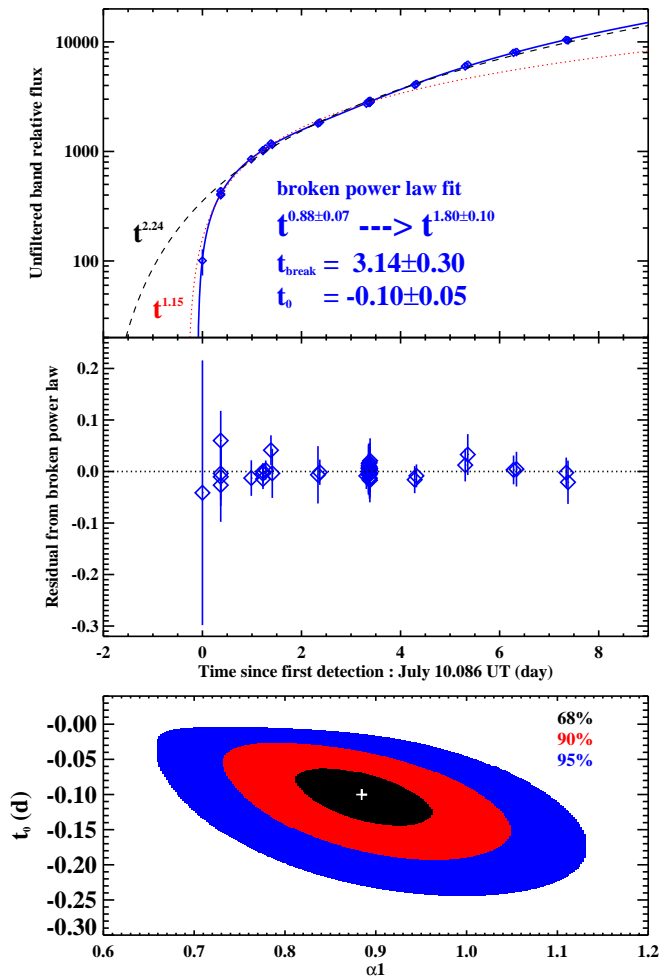}
\caption{Top: the broken power law fit (solid blue) to the early
  unfiltered light curve, with the residuals shown in the middle panel.
  Compared with the two single power law results, the broken power law
  clearly improves the fit; the power law index changes from 0.88 to 1.80, 
  with a break time of 3.14\,d and a first-light time of $-0.10$\,d; see 
  text for details. Bottom: map of the $\chi^2$ hypersurface 
  around the minimum-fit result of $t_0$ and $\alpha_1$. The outbound for 
  each color of black, red, and blue corresponds to 68\%, 90\%, and 95\% 
  confidence intervals (from inside to outside), respectively. 
  \label{figure_lc_brokenPL}}
\end{figure}

In order to determine the time of first light\footnote{Throughout this
  paper, we refer to the time of first light instead of explosion
  time, since the SN may exhibit a ``dark phase" which can last for a
  few hours to days, between the moment of explosion and the first
  observed light (e.g., Rabinak, Livne, \& Waxman 2012; Piro \& Nakar
  2012, 2013).}, one can assume that the SN luminosity scales as the
surface area of the expanding fireball, and therefore increases
quadratically with time ($L \propto t^{2}$, commonly known as the
$t^2$ model; Arnett 1982; Riess et al. 1999; Nugent et al. 2011).
We restrict our model fit to the unfiltered data, which have the best phase
coverage. Although ISSP images are also unfiltered,
there might be possible differences between the KAIT and ISSP 
effective bandpasses.
Fortunately, the second and third ISSP observations are between KAIT observations, 
and the ISSP magnitudes are consistent
with the KAIT light curve, suggesting that the ISSP unfiltered band is very close to that of KAIT.
Moreover, we measured isolated reference stars in the ISSP images and compared their
magnitudes with the same stars in the KAIT images,
finding consistent results between the two telescopes with differences $<0.04$\,mag.
Thus, it is reasonable to combine the ISSP and KAIT unfiltered results.

Regardless, we first apply the fit only to KAIT fluxes in the first few
days (before July 18).
We find that a $t^2$ model cannot fit the data very well.
We therefore free the exponent of the power law and obtain a best-fit value of $2.24\pm0.08$,
with a corresponding first-light time of $-2.14\pm0.17$\,d (relative to the first detection time, July~10.086).
The exponent is about 3$\sigma$ away from the $t^2$ model (marginally consistent).
However,
as can  be seen from the residual plot in Figure \ref{Fig_lcfit}, the first
night of KAIT data is below the fit, indicating an even faster light curve.
This becomes more drastic if we include the first ISSP detection, which
is far below the extrapolation of the $t^{2.24}$ fit. Thus, we refit the fluxes
including both ISSP and KAIT data, but restricted to data taken before July 14.
We find the best-fit power law exponent for these early data
to be $1.15\pm0.04$, with a corresponding first-light time of $-0.31\pm0.05$\,d.
Note that the nondetection from KAIT on July 8.47 (limiting magnitude $\sim19.4$)
is consistent with both the $t^{1.15}$ fit and the $t^{2.24}$ fit.

The apparent change of the power law indices indicates a varying 
power law of the early rising
light curve. Hence, we adopt a broken power law function, also widely
used for fitting GRB afterglows (e.g., Zheng et al. 2012):
\begin{equation}
f = \left(\frac{t-t_0}{t_b}\right)^{\alpha_1} \Big{[} 1 +
\left(\frac{t-t_0}{t_b}\right)^{s({\alpha}1-{\alpha}2)}\Big{]}^{-1/s},
\end{equation}
where $f$ is the flux, $t_0$ is the first-light time, $t_b$ is the break 
time, ${\alpha}1$ and ${\alpha}2$ are the two power law indices before 
and after the
break, and $s$ is a smoothing parameter. The final fit result gives
$t_0=-0.10\pm0.05$\,d,
%(1$\sigma$ error; 3$\sigma = ^{+0.08}_{-0.15}$), 
namely July
9.99, and $t_b=3.14\pm0.30$\,d, $s=-6.32\pm3.26$, ${\alpha}1=0.88\pm0.07$, and
${\alpha}2=1.80\pm0.10$, as shown in Figure \ref{figure_lc_brokenPL}.

With an estimated first-light time of $-0.10$\,d (2.4\,hr), 
this is the earliest detection of any SNe~Ia, even earlier than 
for SN~2011fe (detected
only 11.0\,hr after first light; Nugent et al. 2011) and SN 2009ig 
(detected 17\,hr after first light; Foley et al. 2012).
It also makes SN~2013dy a rare case with more than one detection within 
the initial day after first light: 
there are 3 epochs of detection within 1\,d and 5 epochs within 1.5\,d.
%A caveat for the above conclusion is that our estimate of the first-light 
%time is the direct fitting result from the broken power law, while the 
%actual SN explosion time may be hours to days before first light, and 
%hence may deviate from $-0.1$\,d with larger uncertainty.

Our best-fit broken power law model of the early light curve yields 
the following conclusions.
(1) The $t^2$ model is not sufficient for every SN~Ia; some SNe may have different power law exponents describing
their rise (see also Piro \& Nakar 2012).
(2) The rising exponent may vary with time.
Perhaps the usual $t^2$ model works well for previous SNe~Ia 
because those examples
did not have more than one observation to constrain
the power law exponent within the first day.
The varying exponent indicates that the very early fireball may
exhibit significant changes in either the photospheric temperature, the velocity, or the fireball input energy
during expansion. These changes may happen on a time scale of 2--4\,d after 
first light.
The very early light curve before the break time may be the 
contribution from the shock-heated cooling emission after shock 
breakout, which has a predicted rising index of 1.5 
($f \propto t^{1.5}$; see Eq. 3 in Piro \& Nakar 2013). However, 
our observed power law index is 0.88, smaller than predicted. 
The rising index also depends on underlying physical parameters; 
detailed analysis will be presented elsewhere.

Alternatively, the early-time observations constrain
the emission from the ejecta, which can be used to limit the
radius of the progenitor star as well as interaction with the 
circumstellar medium or a companion star (Kasen 2010). 
For SN~2013dy, the early ISSP unfiltered observation of 
$\sim17.89$ mag (corrected for extinction) at 0.10\,d
limits any emission from this process to be 
$\nu L_{\nu}\lesssim2.6\times10^{40}$\,erg\,s$^{-1}$ at 
optical wavelengths.
Comparing these parameters with those of SN~2011fe, which has 
a constraint on its progenitor star $R_0\lesssim0.1\,{\rm R}_\sun$ 
(see Fig. 4 of Nugent et al. 2011),
our constraint for SN~2013dy is slightly weaker (factor of $\sim2.6$),
and so we infer the radius of the
progenitor star to be $R_0\lesssim0.25\,{\rm R}_\sun$. Even if we 
conservatively assume the first-light time to be
earlier, the same time as the KAIT upper limit (July 8.47), we can still 
find that $R_0\lesssim0.35\,{\rm R}_\sun$,
consistent with a white dwarf progenitor.

\subsection{Spectra}\label{ss:spectra}

\begin{figure}[!hbp]
\centering
\includegraphics[width=.49\textwidth]{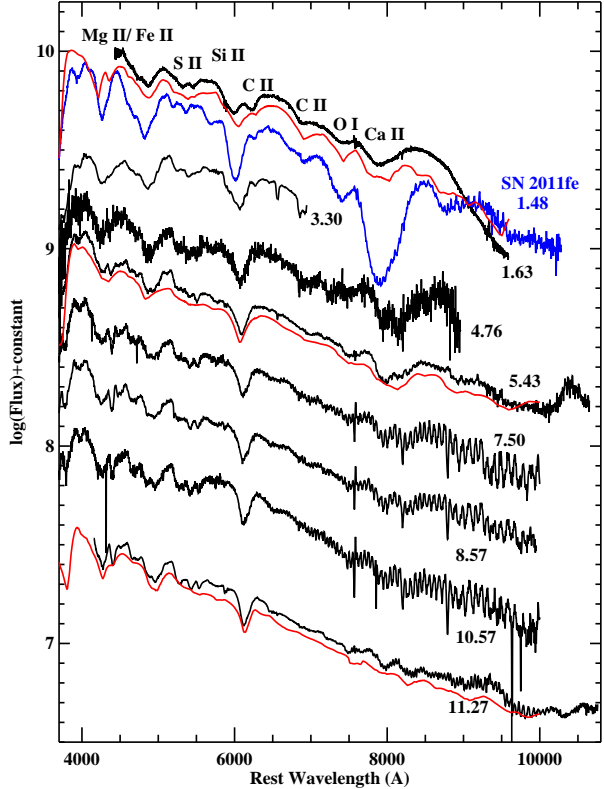}
\caption{Spectra of SN~2013dy
  and a few SYNAPPS fits (red),
  along with comparison to the young SN~2011fe (blue).
  Each spectrum is labeled with its age relative to first light.
  Some major spectral features are labeled at the top.
  Wiggles redward of $\sim7500$\,\AA\ in some of the spectra are
  produced by CCD fringing.}
\label{Fig_spec_all}
\end{figure}

Figure~\ref{Fig_spec_all} shows our spectra of SN~2013dy from the first
\about2\,weeks. Most exhibit narrow \ion{Na}{1}~D
absorption from both the host galaxy and the Milky Way.
The median redshift determined from
these features is $z=0.00383\pm0.00025$, consistent with the
redshift given in SIMBAD (0.00389).

%\begin{figure}[!]

The equivalent width (EW) of \ion{Na}{1}~D absorption is often
converted into reddening, but with large scatter over the
empirical relationship (Poznanski et al. 2011).
The median EW of \ion{Na}{1}~D from the 
host galaxy is measured to be \about0.53\,\AA, which yields a range of
possible reddening values around
$E(B-V)_\textrm{host} = 0.15$\,mag (Poznanski et al. 2011).
For Milky Way extinction, the measured median EW of \ion{Na}{1}~D is \about0.50\,\AA,
corresponding to $E(B-V)_\textrm{MW} = 0.14$\,mag, consistent
with the value of $E(B-V)_\textrm{MW} = 0.15$\,mag given by 
Schlegel et al. (1998); here we adopt the latter.

\subsubsection{Species and Individual Lines}\label{sss:syn}

\begin{figure}[!hbp]
\centering
\includegraphics[width=.49\textwidth]{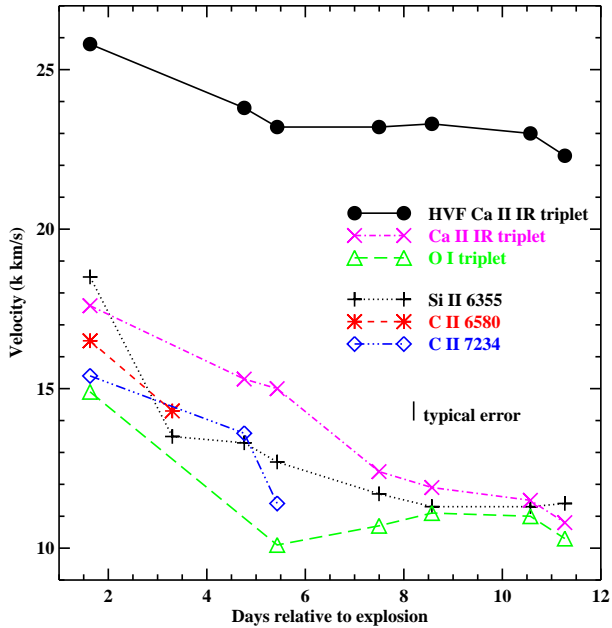}
\caption{Expansion-velocity evolution of different lines measured from the spectra
  of SN~2013dy.
  Uncertainties are $\sim300$\,km\,s$^{-1}$, and are comparable to
  the size of the data points.}
\label{Fig_spec_line_velocity}
\end{figure}

To help identify the species present in our spectra of SN~2013dy, we
used the spectrum-synthesis code {\tt SYNAPPS} (Thomas et al. 2011).
A few examples of our fits are shown in Figure~\ref{Fig_spec_all}.
Our first spectrum of
SN~2013dy (1.63\,d after first light) consists of absorption features from
ions usually seen in SNe~Ia (\ion{Ca}{2}, \ion{Si}{2},
\ion{Fe}{2}, \ion{S}{2}, and \ion{O}{1}, as well as
%\ion{Mg}{2},
strong \ion{C}{2}).
All of these species have expansion velocities $\ga$\,15,000\,\kms, similar to what
was found in the earliest spectra of SN~2011fe (Parrent et al. 2012). Figure
\ref{Fig_spec_line_velocity} shows our measurements of individual line velocities
(see Silverman et al. 2012c for details).

%\begin{figure}[!]

In addition to the usual photospheric absorption component of the \ion{Ca}{2} near-infrared triplet,
SN~2013dy exhibits a high-velocity feature (HVF) in our early
spectra having a velocity of $\sim$\,26,000\,\kms. Similar absorption is also seen in
a few other well-observed SNe, including SN~2005cf (Wang et al. 2009) and
SN~2012fr (e.g., Maund et al. 2013; Childress et al. 2013).
This HVF appears to be detached from the
rest of the photosphere, slowing down to $\sim$\,23,000\,\kms after three days (measured from the first
spectrum) and maintaining that velocity through at least 11\,d.
As for \ion{Si}{2} $\lambda$6355,
the velocity continuously slow down from $\sim$\,18,500\,\kms at 1.63\,d to
$\sim$\,11,400\kms\, at 11.27\,d.

Interestingly, our first spectrum exhibits a strong 
line $\sim245$\,\AA\ redward of the usual prominent \ion{Si}{2} $\lambda$6355.
It is very likely to be the \ion{C}{2} $\lambda$6580 line; a weaker 
\ion{C}{2} $\lambda$7234 feature is also visible.
Such strong \ion{C}{2} lines are not usually seen
in normal SNe~Ia (Silverman et al. 2012b), but similar features 
have been observed in a few super-Chandrasekhar mass examples.
Though \ion{C}{2} is distinguishably detected
in over 1/4 of all normal SNe~Ia (e.g., Parrent et al. 2011; Silverman et al. 2012b), 
it is usually not very strong. However, spectra of other SNe~Ia
have generally not been obtained as early as our spectra of SN~2013dy. 
In fact, the \ion{C}{2} $\lambda$6580 line weakens rapidly in 
SN~2013dy; it became much weaker by 3.30\,d, and it is undetectable 
after an age of $\sim1$ week.
Thus, the early discovery of SNe~Ia and timely spectroscopic 
observations are crucial for detecting the \ion{C}{2} features and 
studying their evolution.

The velocity of \ion{C}{2} $\lambda$7234 is
slightly lower than that of \ion{C}{2} $\lambda$6580 in the 1.63\,d spectrum,
and both are also a bit below that of the photospheric component 
of \ion{Si}{2} $\lambda$6355, as seen in previous work (e.g., Silverman et al. 2012b).
But after $\sim3$\,d, their velocities are similar to each other.
The presence of \ion{C}{2} with velocity comparable to that of \ion{Si}{2} gives direct
evidence that there exists some amount of unburned material. 
Moreover, the presence of both \ion{O}{1}
(often seen in normal SNe~Ia) and \ion{C}{2} suggests that the progenitor is
probably a C+O white dwarf, consistent with the analysis of our early-time light curve.

\subsubsection{Classification}\label{sss:classification}

Using the SuperNova IDentification code (SNID; Blondin \& Tonry 2007), we find that SN~2013dy is
spectroscopically similar to several normal SNe~Ia, though some of our early spectra (7.50, 8.57, 10.57\,d)
also resemble those of the peculiar SN~1999aa and similar events (e.g., Li et al. 2001).
Since the peak $B$-band brightness lies in the range
of typical SN~Ia luminosities, SN~2013dy is probably a normal SN~Ia.

%%%%%%%%%%%%%%%%%%%%%%%%%%%
%%  Section 4:  Conclusion  %%
%%%%%%%%%%%%%%%%%%%%%%%%%%%

\section{Conclusions}\label{s:conclusions}

In this {\it Letter} we present optical photometry and spectroscopy
of the Type Ia SN~2013dy,
the earliest detection of an SN~Ia thus far.
The rising light curve shows a variable power-law exponent and its early-time spectrum exhibits
a strong \ion{C}{2} feature, both of which are not seen in previous studies of normal SNe~Ia.
Such well-studied objects will help us 
understand the underlying nature of SNe~Ia.

\begin{acknowledgments}

A.V.F.'s group (and KAIT) at UC Berkeley have received
financial assistance from the TABASGO Foundation, the Sylvia \& Jim
Katzman Foundation, the Christopher R. Redlich Fund, and NSF grant
AST-1211916.  J.M.S. is supported by an NSF
postdoctoral fellowship under award AST-1302771.  X. Wang acknowledges
NNSFC grants
11073013 and 11178003, the Foundation of Tsinghua University
(2011Z02170), and the Major State Basic Research Development Program
(2013CB834903). J.V. is grateful for Hungarian OTKA grant NN 107637.
J.C.W. acknowledge support from NSF AST-1109801.
This research used resources of NERSC, supported by 
DoE under Contract DE-AC02-05CH11231.
Some data were obtained at the W.~M. Keck 
Observatory, which was made possible by the generous financial support of 
the W.~M. Keck
Foundation. We thank the staffs of the various observatories at which
data were obtained.
We thank the anonymous referee for the useful suggestions that improved the paper.

\end{acknowledgments}

\end{document}